\documentclass[fleqn,usenatbib]{mnras}

\usepackage{newtxtext,newtxmath}
\usepackage{xcolor}

\usepackage[T1]{fontenc}

\usepackage{soul}

\DeclareRobustCommand{\VAN}[3]{#2}
\let\VANthebibliography\thebibliography
\def\thebibliography{\DeclareRobustCommand{\VAN}[3]{##3}\VANthebibliography}

\usepackage{graphicx}	
\usepackage{amsmath}	
\usepackage[version=4]{mhchem} 
\usepackage{gensymb}
\usepackage{upgreek}

\title[Far-infrared spectroscopy of photostable PAHs]{Gas-phase spectroscopy of photostable PAH ions from the mid- to far-infrared}

\author[S. D. Wiersma et al.]{
Sandra D. Wiersma,$^{1,2,3}$
Alessandra Candian,$^{1}$
Joost M. Bakker$^{2}$
and Annemieke Petrignani$^{1}$\thanks{E-mail: a.petrignani@uva.nl (UvA)}
\\
$^{1}$Van `t Hoff Institute for Molecular Sciences, University of Amsterdam, PO Box 94157, 1090 GD, Amsterdam, The Netherlands\\
$^{2}$Radboud University, Institute for Molecules and Materials, FELIX Laboratory, Toernooiveld 7, 6525 ED Nijmegen, The Netherlands\\
$^{3}$Institut de Recherche en Astrophysique et Planétologie (IRAP), CNRS, Université de Toulouse (UPS), 31028 Toulouse, France
}

\date{Accepted XXX. Received YYY; in original form ZZZ}

\pubyear{2020}

\begin{document}
\label{firstpage}
\pagerange{\pageref{firstpage}--\pageref{lastpage}}
\maketitle
\newcommand{\cm}{cm$^{-1}$}
\begin{abstract}{
We present gas-phase InfraRed Multiple Photon Dissociation (IRMPD) spectroscopy of cationic phenanthrene, pyrene, and perylene over the 100--1700 cm$^{-1}$ (6--95 $\upmu$m) spectral range. This range covers both local vibrational modes involving C--C and C--H bonds in the mid-IR, and large-amplitude skeletal modes in the far-IR. The experiments were done using the 7T Fourier-Transform Ion Cyclotron Resonance (FTICR) mass spectrometer integrated in the Free-Electron Laser for Intra-Cavity Experiments (FELICE), and findings were complemented with Density Functional Theory (DFT) calculated harmonic and anharmonic spectra, matching the experimental spectra well. The experimental configuration that enables this sensitive spectroscopy of the strongly-bound, photo-resistant Polycyclic Aromatic Hydrocarbons (PAHs) over a wide range can provide such high photon densities that even combination modes with calculated intensities as low as 0.01 km$\cdot$mol$^{-1}$ near 400 cm$^{-1}$ (25 $\upmu$m) can be detected. Experimental frequencies from this work and all currently available IRMPD spectra for PAH cations were compared to theoretical frequencies from the NASA Ames PAH IR Spectroscopic Database to verify predicted trends for far-IR vibrational modes depending on PAH shape and size, and only a relatively small redshift (6--11 cm$^{-1}$) was found between experiment and theory. The absence of spectral congestion and the drastic reduction in bandwidth with respect to the mid-IR make the far-IR fingerprints viable candidates for theoretical benchmarking, which can aid in the search for individual large PAHs in the interstellar medium.
}
\end{abstract}





\section{Introduction}

Polycyclic aromatic hydrocarbons (PAHs) are one of the most stable families of organic molecules known. They are found in outer space, where they are formed through combustion-like chemistry in the outflows of carbon-rich stars \citep{Allamandola1989,Latter1991}. The presence of PAHs has been established in many astronomical environments, and it has been estimated that PAHs comprise between 10--18\% of the elemental carbon budget \citep{Joblin1992, Tielens2013}. 
Since the interstellar PAH hypothesis was presented in 1984  \citep{Sellgren1984, Leger1984}, many astronomical observations, theoretical studies and laboratory experiments have been conducted, that have led to the conclusion that the dominant Aromatic Infrared Bands (AIBs) observed in the mid-infrared (MIR) at 3.3, 6.2, 7.7 and 11.3 $\upmu$m, can be attributed to emissions of UV-excited PAHs in the ISM \citep{Hudgins1995, Allamandola1989,VanDishoeck2004,VanDiedenhoven2004, Tielens2008,Peeters2011,Montillaud2013,Candian2014,Croiset2016,Andrews2016,Bauschlicher2018}. These and many other studies have helped put limits on the expected size of interstellar PAHs and elucidated details of their life cycles; it is now well-established that a large fraction of the PAH population in the ISM is present in cationic form (see \cite{Tielens2008} and references therein).

Recently, the radio wave fingerprints of both pure and nitrogen-containing PAHs with two rings have been detected in the dark cloud TMC-1 \citep{McGuire2021,Cernicharo2021,Burkhardt2021}. Fingerprints of closely related species have also been detected, leading to the discovery of benzene and benzonitrile \citep{Cernicharo2001, Kraemer2006, McGuire2018}, as well as the fullerenes \ce{C60}, \ce{C60+} and \ce{C70} \citep{Cami2010, Sellgren2010, Berne2013, Campbell2015, Cordiner2019}. The presence of fullerenes has been confirmed by both their visible \citep{Campbell2015,Cordiner2019} and MIR \citep{Cami2010,Sellgren2010,Berne2013} fingerprints.
Benzene, \ce{C60}, and \ce{C70} possess characteristic IR emission lines that allowed interstellar MIR features to be assigned to them. However, an infrared match with the spectrum of a single PAH, especially for a large one ($>50$ C atoms) suspected to be abundant \citep{Andrews2016} is still elusive, as PAHs in the MIR exhibit highly similar spectra, dominated by local \ce{C-C} and \ce{C-H} modes.

\begin{figure}
\centering
\includegraphics[width=\linewidth]{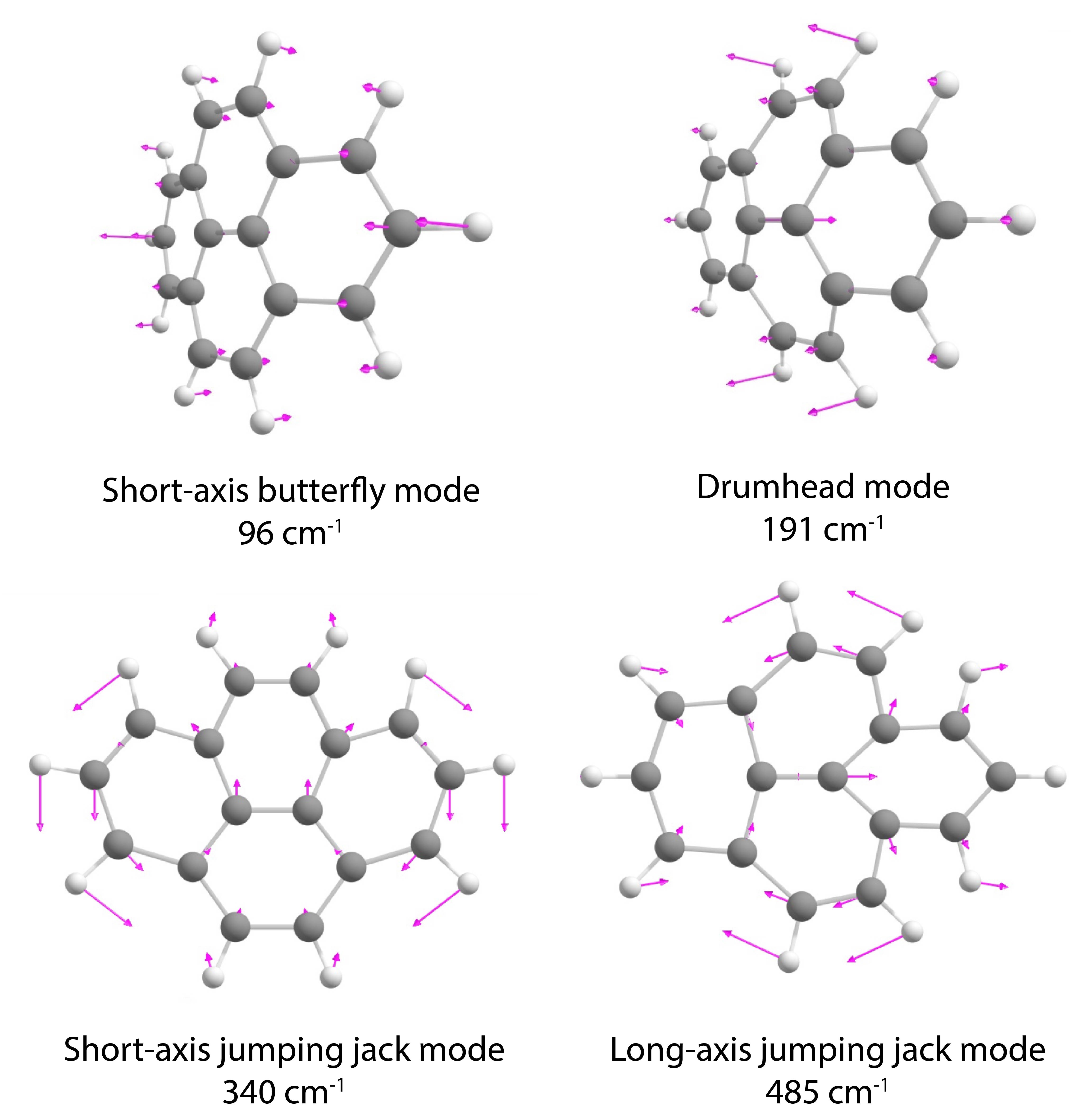}
\caption{\label{fig:modes} Skeletal vibrations of cationic pyrene, visualised with displacement vectors. The scaled frequencies are taken from the  harmonic DFT calculations which are treated in Section 2.2 and listed from the drumhead mode on in Table \ref{tab:pyr}.}
\end{figure}

The scenario is different for the far-infrared (FIR) fingerprints of PAHs. Here, global, skeletal deformation vibrations, with names such as drumhead, butterfly or jumping jack modes (see Fig. \ref{fig:modes}) are much more molecule specific \citep{Ricca2010, Ricca2012, Mulas2006a, Mulas2006b}. Although possible FIR PAH emission bands are often obscured by dust emissions in interstellar objects \citep{Draine2003}, emission features in the MIR-to-FIR transition zone, at 15.8, 16.4, 17.4, 17.8, and 18.9 $\upmu$m, have been attributed to PAHs \citep{Boersma2010, Ricca2010}. Further in the FIR, bands have been suggested to be even more characteristic for individual PAHs or PAH families \citep{Mulas2006a, Mulas2006b, Joblin2011, Boersma2011a, Boersma2011b,Ricca2012}. These studies are, however, entirely based on theoretical spectra from the NASA Ames PAH database \citep{Bauschlicher2018}, for which the FIR benchmarking is primarily based on the spectrum of a single PAH, neutral naphthalene \citep{Pirali2009}.

Laboratory studies have revealed many characteristic IR spectral properties of PAHs, but have mostly been limited to wavelengths below 20 $\upmu$m (above 500 cm$^{-1}$), which do not involve the truly global modes. Matrix isolation spectroscopy studies have mostly focussed on neutral PAH species, partially due to difficulties in discriminating signal originating from ions from those from absorption by neutral species \citep{Leger1989,Blanco1990, Moutou1995,Mattioda2009,Cataldo2013,Zhang2017}. Other difficulties including impurities and backgrounds exist for gas-phase FT-IR or emission spectroscopy \citep{Kurtz1992,Zhang1996,Cane1997, Pirali2006,Pirali2009, Pirali2013,Goubet2014}. 
For neutral PAHs, ion-dip spectroscopy allows for detailed high-resolution investigations of, \textit{e.g.}, anharmonic behaviour; however, its application to ionic species is not trivial. \citep{Maltseva2015,Lemmens2019}.
To record laboratory spectra for gas-phase ions over large spectral ranges (covering mid- to far-IR), IR multiple-photon dissociation (IRMPD) spectroscopy is presently one of the most powerful methods \citep{Oomens2000, Oomens2006}. In the FIR, this type of action spectroscopy is, however, far from trivial due to the requirement that several tens to hundreds of IR photons need to be absorbed to overcome the fragmentation barrier in strongly bound species such as PAHs. With both decreasing IR absorption cross-sections and the need to absorb more photons at longer wavelengths, laboratory spectroscopy studies of gas-phase PAHs have thus far been mostly limited to the MIR 3--18 $\upmu$m range \citep{Piest1999, Oomens2000, Piest2001, Bakker2011, Bouwman2019, Lemmens2019b, Palotas2020}. 

In this work, we used the high IR photon densities available in the Fourier Transform Ion Cyclotron Resonance (FT-ICR) mass spectrometer beamline of the FELICE intra-cavity free-electron laser \citep{Petrignani2016} to perform IRMPD spectroscopy of PAH cations in the FIR up to wavelengths of 100 $\upmu$m. We present IRMPD spectra for phenanthrene (\ce{C14H10+}), pyrene (\ce{C16H10+}), and perylene (\ce{C20H12+}), in the range of 100--1700 cm$^{-1}$ with particular attention to the 100--600 cm$^{-1}$ FIR spectral range. We assign a large range of low-frequency bands using Density Functional Theory (DFT) calculations in both the harmonic and the anharmonic approximation. The narrow FIR modes in our experimental spectra are then considered as possible benchmarking data to improve the NASA Ames PAH Database in this spectral range \citep{Bauschlicher2018}. Finally, we discuss the implications of our findings for possible astronomical detection of molecule-specific signatures in the ISM.

\section{Methods}
\begin{figure*}
\includegraphics[width=0.75\linewidth]{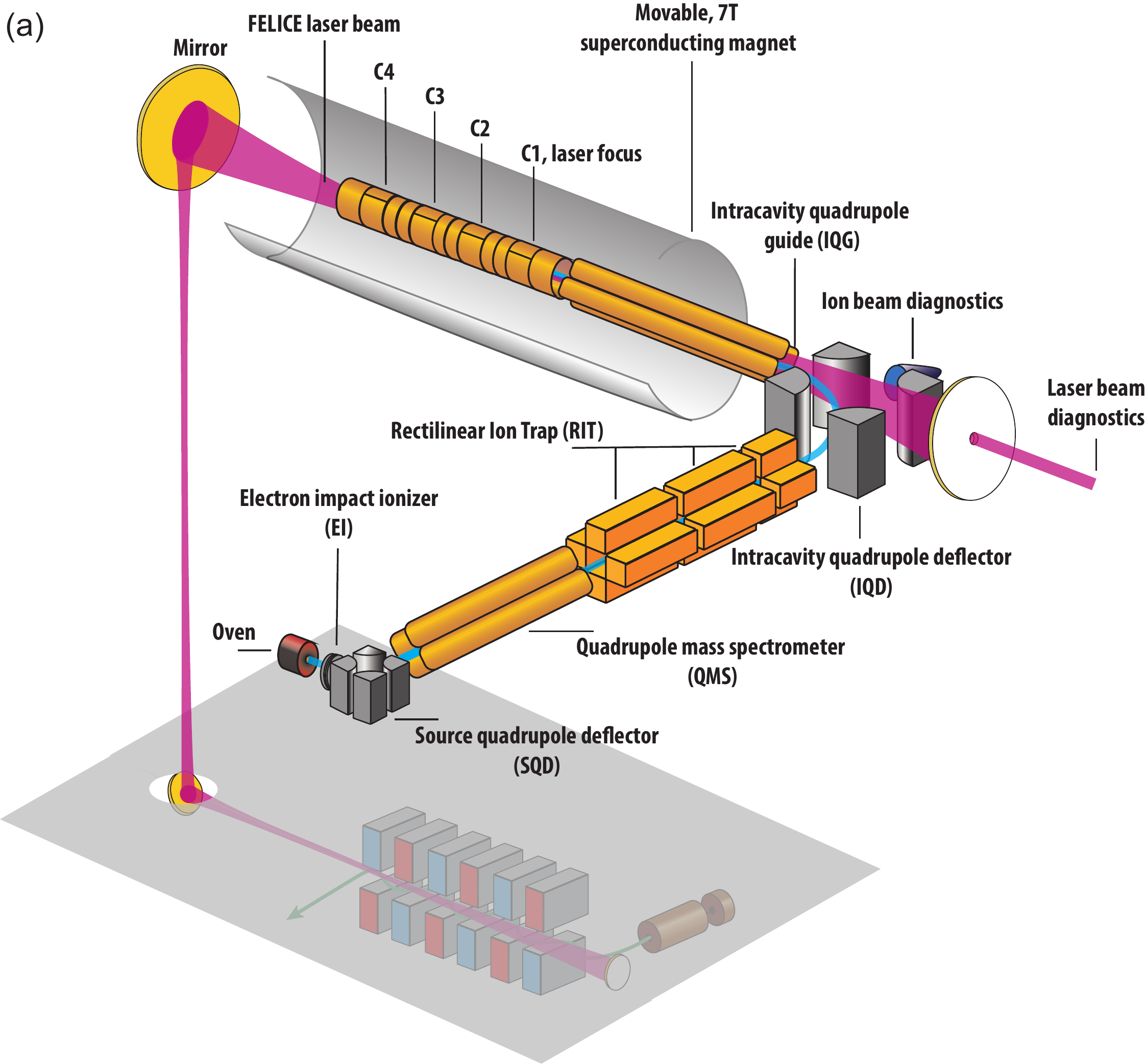}
\newline
\includegraphics[width=0.8\linewidth]{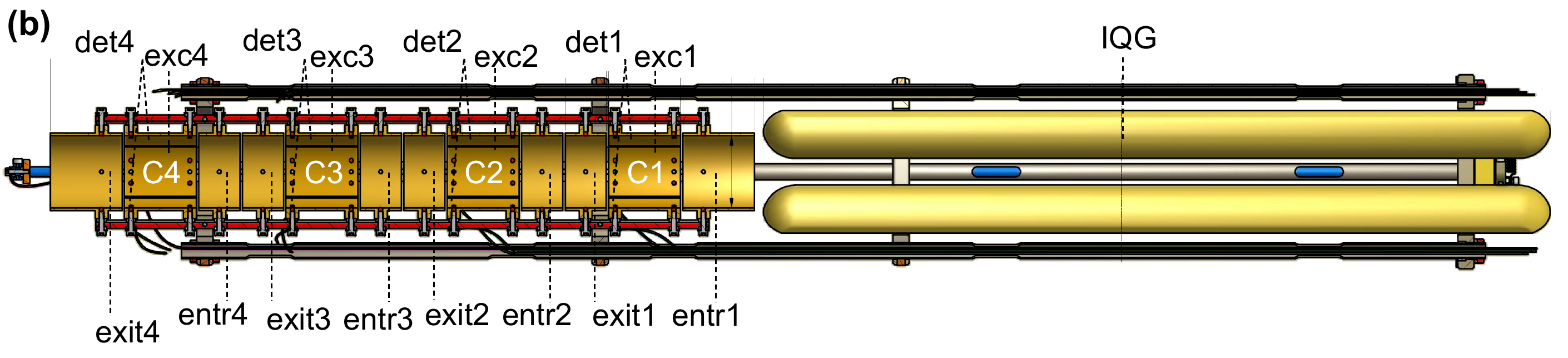}
\caption{\label{fig:setup} (a) Schematic of the FELICE FT-ICR instrument. The path of the molecules through the apparatus from the sublimation source down to the ICR cells is depicted in blue. The fuchsia beam illustrates the FELICE waist through the cavity. Proportions in this figure are not to scale.
(b) Cross section of the four storage cells (C1--4) and the intra-cavity quadrupole ion guide (IQG). The ions move from right to left and are stored in one of the storage cells (C\#). The top and side plates are used as excitation (exc\#) and detection (det\#) plates, respectively. When using one cell, the electrodes of all other cells are grounded. The focus of the FELICE laser beam coincides with the centre of C1, and is characterised by a 82 mm Rayleigh length.}
\end{figure*}

\subsection{Experimental}
The IRMPD spectra were recorded using the Fourier Transform Ion Cyclotron Resonance (FT-ICR) mass spectrometer beam line of the Free Electron Laser for Intra-Cavity Experiments (FELICE), at the FELIX Laboratory in Nijmegen. The FELICE FT-ICR apparatus, schematically shown in Figure \ref{fig:setup}a, has been described in detail before \citep{Petrignani2016,Wensink2020}. Here, additional experimental details are presented on features of the FELICE FT-ICR that were used for the first time in this work.

Phenanthrene, pyrene, and perylene (Sigma Aldrich/Merck) were brought into the gas phase using an effusive sublimation source, at temperatures of 50, 70, and 130\degree C respectively. The gas-phase PAHs were then exposed to 20-eV electrons in an electron impact ioniser (Ardara Technologies), creating singly-charged radical cations. These cations were led through a series of ion optics, including a linear quadrupole mass spectrometer (QMS) operated in radio-frequency guiding mode and collected a sectioned, quadrupole ion trap with rectilinear rods (RIT), and collisionally cooled with argon gas ($\sim 10^{-2}$~mbar) \citep{Ouyang2004}. After accumulating and cooling over 50 ms in the RIT, an ion pulse was extracted towards an intra-cavity quadrupole deflector (IQD), which diverted the ion trajectory by 90\degree\ into the FELICE cavity. A home-built intra-cavity radio-frequency quadrupole ion guide (IQG) transported the ions toward one of the four FT-ICR storage cells. For each cell, the mass resolution exceeds m/$\Delta$m=10$^5$ for mass-to-charge ratios of up to at least 1000 m/z.

The FT-ICR storage cells are labelled C1--C4, and are depicted in Figure \ref{fig:setup}b. These open-ended, cylindrical cells (5 cm diameter) are located within a movable 7-Tesla superconducting magnet (Cryomagnetics Inc.). This magnet can be positioned in such a way that the volume of homogeneous magnetic field ($<$19 ppm) overlaps with the centre of either of the four cells. The multi-cell configuration was introduced to enable IR spectroscopy with varying IR fluences inside the laser cavity. C1 is positioned in the laser focus, and ions stored here experience the largest laser fluence. With a spacing between the cells of 10 cm, and the laser Rayleigh range of 82 mm, ions in each next cell are submitted to a fluence reduced by a factor of 2.3. For compounds with a relatively strong absorption efficiency, spectroscopy is preferably done in C4, as the overlap between the ion cloud and the laser beam is larger, and the reduced fluence will lead to less spectral broadening. C1 is used for the most photostable species or weakest absorptions. For pyrene and perylene, spectra were recorded in C4 in the 600--1700 \cm\ spectral range, and spectra at lower frequencies in C1. For phenanthrene, the 235--650 \cm\ spectral range was recorded using C4, and the 105--300 \cm\ range in C1.

After transfer to the selected cell, the cations of interest were mass-isolated via a stored waveform inverse Fourier Transform (SWIFT) pulse \citep{Marshall1998}. The ions were then exposed to one or several (max. 20) FELICE macropulses. For each spectrum, the IR frequency was scanned in steps of 1, 2.5 or 5 \cm, depending on the spectral range. At each frequency step, 5--20 time transients were averaged and Fourier-transformed. The whole experimental sequence is controlled by home-built software developed by \citet{Mize2004}.

The IRMPD spectrum was obtained by monitoring the fragmentation yield $Y(\nu)$ at wavenumber $\nu$, defined as
\begin{equation}
    Y(\nu)=\ln{\left(\frac{N_\text{par}(\nu)}{N_\text{par}\left(\nu\right)+N_\text{frag}\left(\nu\right)}\right)}\label{eqn}
\end{equation}
with $N_\text{frag}$ and $N_\text{par}$ as the total fragment and parent mass counts, respectively. $Y$ is then divided by $P(\nu)$, the macropulse energy, which is inferred from coupling a fraction of the light out through a 1 mm-diameter hole of the cavity end mirror, and directing it onto a power meter. The IR wavelength was calibrated using a grating spectrometer (Princeton Instruments SpectraPro). 
The 100--1700 cm$^{-1}$ spectral range was recorded with three separate FEL settings. In the 600--1800 cm$^{-1}$ range, the FELICE macropulse energy peaked between 1500--1600 cm$^{-1}$ at 1.6 J. In the 200--600 cm$^{-1}$ range, it was slightly lower with its maximum at 1.3 J around 475 cm$^{-1}$. For the 100--300 cm$^{-1}$ range, the maximum was reached around 275 cm$^{-1}$ at 0.4 J.
After power correction of the three individual curves, they are joined by normalising to the overlapping resonances, or in the case there is no overlapping resonance, by correcting for the drop in fluence from one cell to the next. The composite spectrum is then normalised to the resulting highest intensity, giving the final normalised IRMPD yield.
For the presented experiments, 4--10 $\upmu$s long macropulses were used, at a FELICE repetition rate of 5 or 10 Hz, which in turn consist of ps-long micropulses at 1 GHz. The spectral bandwidth (FWHM) was $\sim0.6$\% of the central wavelength (in $\upmu$m).

\subsection{Computational}
Density functional theory (DFT) calculations was employed to predict the IR spectra and compare to experiment, using the Gaussian16 Suite \citep{g16}. For all three molecules, geometry optimisations and harmonic vibrational calculations were performed using the B3LYP functional \citep{Becke1993, Lee1988} and 6-311++G(2d,p) \citep{Frisch1984} basis set. The harmonic calculations were scaled to match the experimental frequencies by a factor of 0.96. For pyrene and perylene, we present both the scaled harmonic calculations and unscaled anharmonic calculations at the B3LYP/N07D level of theory \citep{Barone2008,Mackie2018}. The individual modes are visualised and identified using the computational chemistry software suite Gabedit \citep{gabedit}. We adopted the FIR mode naming conventions introduced by \citet{Ricca2010}.

\section{Results}

\begin{table*}
\caption{\label{tab:frag} Cationic fragment masses detected in the IRMPD spectroscopy of phenanthrene (\ce{C14H10+}, $m/z=178$), pyrene (\ce{C16H10+}, $m/z=202$) and perylene (\ce{C20H12+}, $m/z=252$). The dominant channel/fragment is denoted in bold. Detected fragment channels that were too weak to be included in the IRMPD spectra are listed in italic.}
\begin{tabular}{clclcl}
\hline
\multicolumn{2}{c}{\ce{C14H10+}}           & \multicolumn{2}{c}{\ce{C16H10+}}   & \multicolumn{2}{c}{\ce{C20H12+}}  \\
\multicolumn{2}{c}{$m/z=$ 178}          & \multicolumn{2}{c}{$m/z=$ 202}  & \multicolumn{2}{c}{$m/z=$ 252} \\
\hline
fragment $m/z$         & loss of     & fragment $m/z$ & loss of          & fragment $m/z$    & loss of  \\
\hline
177                  & \ce{-H}           & 201          & \ce{-H}           & 251             & \ce{-H}       \\
\textbf{176}                  & \textbf{\ce{-2H} }        & \textbf{200 }         & \textbf{\ce{-2H}}          & \textbf{250}             &\textbf{\ce{-2H} }    \\
152                  & \ce{-[2C, 2H]} & 199          & \ce{-3H}         & 249             & \ce{-3H}      \\
151                  & \ce{-[2C, 3H]} & 198          & \ce{-4H}          & 248             & \ce{-4H}      \\
150                  & \ce{-[2C, 4H]} & \textit{176}          & {\it\ce{-[2C, 2H]}} & \textit{246 }            & {\it\ce{-6H}}      \\
\multicolumn{1}{l}{} &               & \textit{174}          & {\it \ce{-[2C, 4H]}} &                 &       \\  
\hline
\end{tabular}
\end{table*}

The fragments observed in the IRMPD spectroscopy of the three respective PAHs are listed in Table \ref{tab:frag}. The smallest PAH ion, phenanthrene, exhibits both H and hydrocarbon loss. For pyrene, H-loss is dominant, and the hydrocarbon loss is very weak. For perylene, only H-loss is observed.  This follows previously established experimental trends for cationic PAHs using IRMPD \citep{Bouwman2019}, collision-induced dissociation \citep{West2019}, and UV-photodissociation \citep{Ekern1998}. 
Theoretical studies on the reaction paths for \ce{[2C, 2H]} loss processes showed that pericondensed PAHs lack the structural flexibility to form the four- and five-ringed intermediates necessary for \ce{[2C, 2H]} loss, and that structural rigidity increases for larger, more aromatically stabilised PAHs \citep{Holm2011,West2019}. Furthermore, the ease with which $-$2H/\ce{-H2} is lost from highly-excited large PAHs is also reflected in other work \citep{Zhen2014,Zhen2017,Zhen2018,Castellanos2018,RodriguezCastillo2018}. 

The IR spectra of the respective PAH ions were constructed according to Eqn. \ref{eqn} using the fragments listed in Table \ref{tab:frag} followed by power correction and normalisation, providing the normalised IRMPD yield. Minor loss channels (italicised) were omitted due to their low S/N ratios.

\subsection{Phenanthrene}

\begin{figure}
\centering
\includegraphics[width=0.8\linewidth]{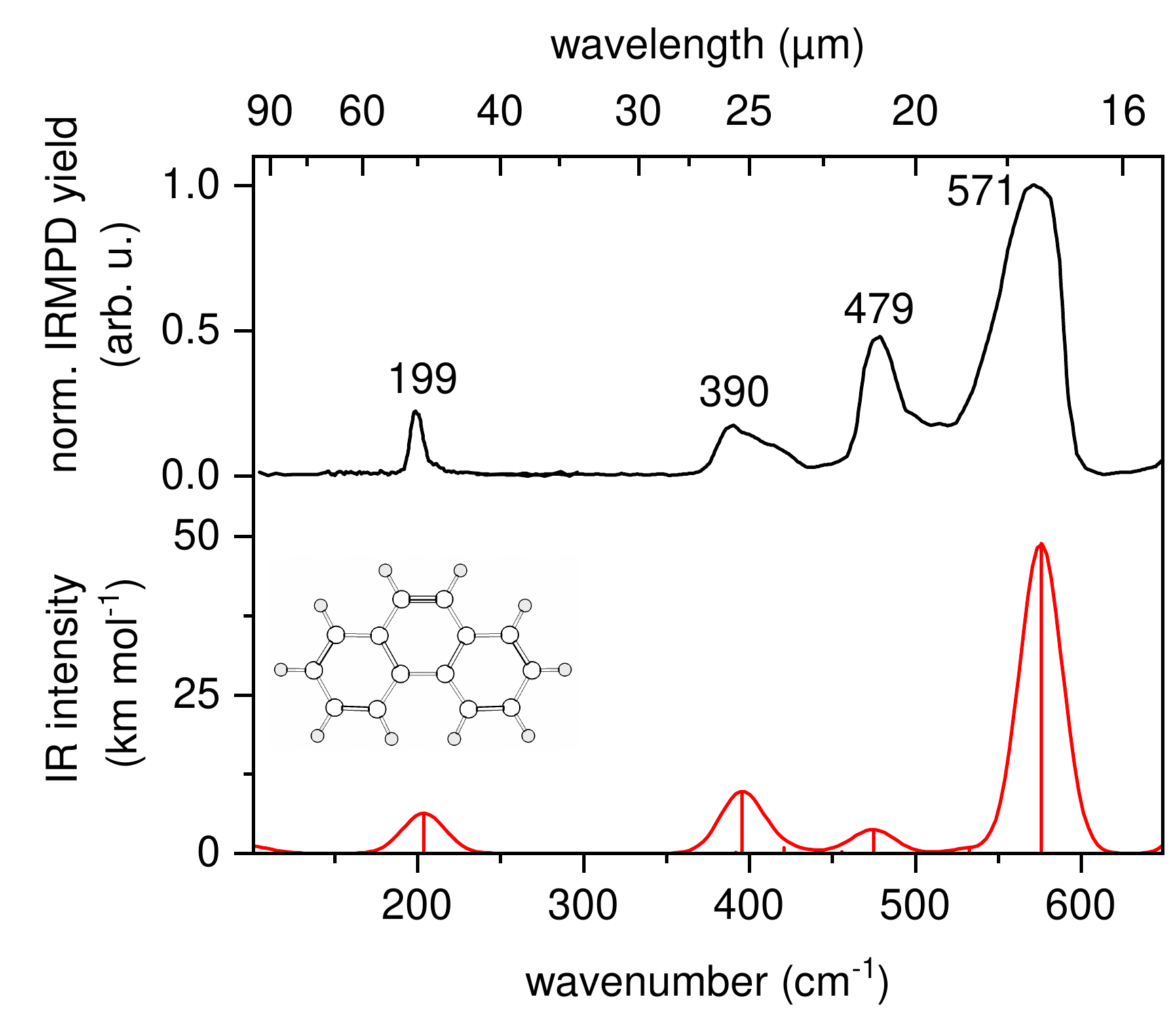}
\caption{\label{fig:phen} The experimental IRMPD spectrum (top) and theoretical IR spectrum (bottom) of cationic phenanthrene. The harmonic theoretical stick spectrum is scaled with a factor 0.96, and convolved with a 30 cm$^{-1}$ FWHM Gaussian line shape function.}
\end{figure}
\newcommand{\km}{km$\cdot$mol$^{-1}$}

\begin{table}

\caption{\label{tab:phen} Phenanthrene cation observed band frequencies (\cm) and relative intensities, together with calculated frequencies (\cm) and IR intensities (\km), and descriptions of the theoretical modes. The listed harmonic frequencies are scaled by a factor 0.96. \textit{oop: out-of-plane; ip: in-plane; E/C: elongation/compression}}
\begin{center}
\begin{tabular}{ccccl}
\hline
\multicolumn{2}{c}{Experiment}&\multicolumn{2}{c}{Harmonic calc.}& Mode description    \\
Freq. & Norm. $Y$ & Freq. & IR int. &                             \\
(\cm) &  & (\cm) & (\km) &                                                  \\
\hline
199 & 0.22 & 204 & 6.3      & long-axis butterfly    \\
390 & 0.18 & 395 & 9.5      & quarto HCCH$_{\text{oop}}$ twist, \\
&&&&  duo CH$_{\text{oop}}$ wag \\
&& 420 & 0.9 & armchair ring rock \\
479 & 0.48 & 475 & 3.7      & ip rock central ring,\\
&&&& E/C outer phenyls \\
 & & 532 & 0.7 & E/C middle ring \\
571 & 1.00 & 576 & 48.9     & E/C outer phenyls\\
\hline
\end{tabular}
\end{center}
\end{table}

The IR spectrum of cationic phenanthrene (\ce{C14H10+}, $m/z=178$) is shown in the 105--650 \cm\ spectral range in Fig. \ref{fig:phen} (black, top). The spectrum is a composite of two scans, which overlap between 240 and 290 \cm. Four well-resolved resonances are found, for which frequencies and intensities are listed in Table \ref{tab:phen}. The bandwidths observed are roughly 20 \cm\ full-width at half-maximum (FWHM), except for the 199 \cm\ band, which has a width of $\sim5$ \cm. We attribute this smaller bandwidth in the largest part to the diminished effects of anharmonicity at lower levels of vibrational excitation \citep{Joblin1994,Chakraborty2021}, and to a reduction of the power broadening at longer wavelengths and the reduced FEL spectral bandwidth. The 390 \cm\ band exhibits a high-frequency shoulder, and the 571 cm$^{-1}$ band is asymmetric, both likely caused by underlying low-intensity resonances.
The shown spectrum partially overlaps the IR spectra reported by \citet{Piest2001}. \citet{Piest2001} presents the IRMPD spectrum of the phenanthrene cation and an IR photodissociation spectrum of the cold phenanthrene-Ar cation down to approximately 400 \cm. For the untagged phenanthrene cation, a band was observed at 578 cm$^{-1}$, while the tagged spectrum reported bands at 482 and 581 cm$^{-1}$.  The same publication also reported multiple Franck-Condon active vibrational states detected via mass-analysed threshold ionisation (MATI) further into the FIR, of which only the band at 402 cm$^{-1}$ matches with this work; the other reported bands are IR-inactive. 
The results from Piest \textit{et al.} are consistent with this work, with a maximum frequency deviation of 10 cm$^{-1}$.

Our DFT-calculated spectrum of phenanthrene is also shown in Fig. \ref{fig:phen} (bottom panel, red) with the calculated frequencies scaled by a factor of 0.96 and listed in Table \ref{tab:phen}. To facilitate comparison with the experiment, the stick spectrum is convolved with a Gaussian line shape function with a 30 \cm\ FWHM. 
One can observe that there is a good agreement between calculated and experimental band positions. The calculated relative intensities compare reasonably well, with the largest mismatch for the 479 \cm\ experimental band, which is predicted far weaker than observed. The high-frequency shoulder of the 390 \cm\ band is rationalised by a (weak) predicted band at 420 \cm, and the raised baseline between the 479 and 571 \cm\ bands is attributed to a calculated band at 532 \cm. A detailed comparison between the experiment and the DFT calculation can be found in Table \ref{tab:phen}, along with descriptions of the vibrational modes. These mode descriptions show that as we move further into the FIR, the skeletal motions become increasingly out-of-plane. 

\subsection{Pyrene}
\begin{figure*}
\includegraphics[width=1.0\textwidth]{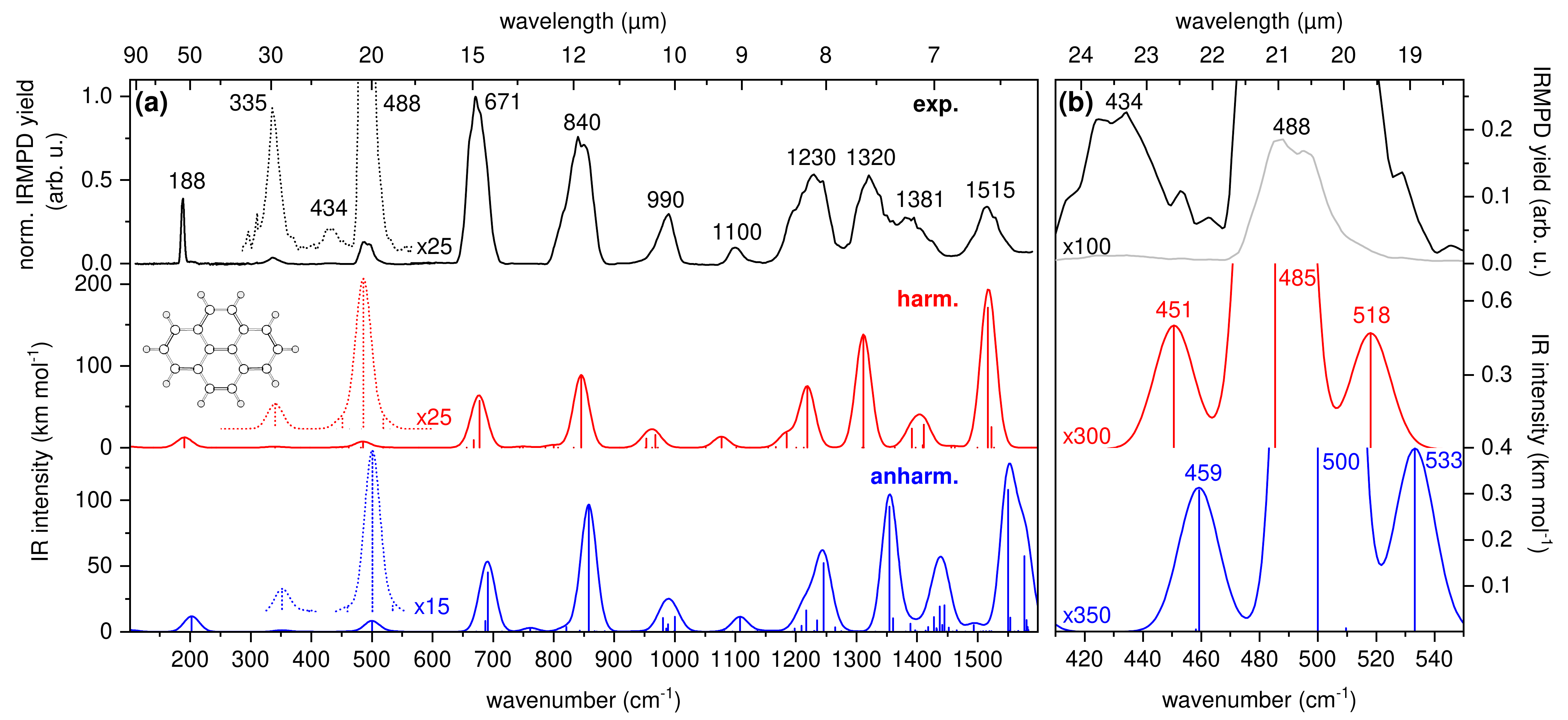}
\caption{\label{fig:pyr} a) The experimental IR spectrum of cationic pyrene (top, black), accompanied by DFT calculated spectra in harmonic (middle/red, frequencies scaled by a factor 0.96) and  anharmonic (bottom/blue, unscaled) approximations; both calculated spectra are convolved with a 30 cm$^{-1}$ FWHM Gaussian line shape function. The dotted curves are vertical enlargements of the solid curves, slightly raised from the baseline for clarity. b) Close-up of 400--550 \cm\ spectral range of panel a), with a convolution of 15 \cm\ for both calculated spectra.}
\end{figure*}

\begin{table*}
\caption{\label{tab:pyr} Pyrene cation observed band frequencies (\cm) and relative intensities, together with calculated frequencies (\cm) and IR intensities (\km), and descriptions of the theoretical modes. The listed harmonic frequencies are scaled by a factor 0.96. \textit{oop: out-of-plane; ip: in-plane; E/C: elongation/compression}}
\begin{tabular}{ccccccl}
\hline
\multicolumn{2}{c}{Experiment}& 
\multicolumn{2}{c}{Harmonic calculations}&
\multicolumn{2}{c}{Anharmonic calculations}&\multicolumn{1}{l}{}\\
Freq. & Norm. $Y$ & Freq. & IR int. & Freq. & IR int. & \multicolumn{1}{l}{Mode description} \\
(cm$^{-1}$) &  & (cm$^{-1}$) & (km$\cdot$mol$^{-1}$) & (cm$^{-1}$) & (km$\cdot$mol$^{-1}$) & \multicolumn{1}{l}{} \\
\hline
188  & 0.390 & 191  & 12.4     & 203    & 11.7     & drumhead   \\
335  & 0.036 & 340  & 1.2      & 352    & 1.2      & short-axis jumping jack  \\
434  & 0.003 & 451  & 0.5      & 459    & 0.3      & long-axis HCCH$_{\text{oop}}$ twist              \\
488  & 0.130 & 485  & 7.3      & 500    & 8.3      & long-axis jumping jack \\
     &       & 518  & 0.5      & 533    & 0.4      & short-axis E/C  \\
671  & 1.000 & 668  & 9.0      & 687    & 8.4      & long-axis E/C \\
     &       & 678  & 57.2     & 691    & 45.3     & in-phase duo and trio CH$_{\text{oop}}$ wag      \\
840  & 0.759 & 845  & 88.7     & 858    & 95.7     & in-phase duo and trio CH$_{\text{oop}}$ wag      \\
990  & 0.298 & 953  & 11.1     & 981    & 10.6     & short-axis E/C\\
     &       & 968  & 15.4     & 1000   & 11.5     & long-axis E/C   \\
1100 & 0.096 & 1077 & 13.4     & 1108   & 11.1     & antisymm. CH$_{\text{ip}}$ trio scissor \\
     &       & 1184 & 17.3     & 1217   & 16.3     &        \\
1230 & 0.533 & 1218 & 74.9     & 1246   & 52.4     & in-phase CH$_{\text{ip}}$ duo rock     \\
1320 & 0.528 & 1311 & 138.3    & 1354   & 95.2     & short-axis ipCC stretch   \\
     &       &      &          & 1360   & 10.2     &                    \\
     &       &      &          & 1389   & 6.3      &                   \\
1381 & 0.276 & 1391 & 23.2     & 1428   & 11.1     & CH$_{\text{ip}}$ scissor, short-axis CC$_{\text{ip}}$ stretch  \\
     &       & 1411 & 28.1     & 1437   & 19.3     & CH$_{\text{ip}}$ scissor, long-axis CC$_{\text{ip}}$ stretch   \\
     &       &      &          & 1445   & 20.3     &                       \\
1515 & 0.341 & 1517 & 171.2    & 1550   & 108.2    & CH$_{\text{ip}}$ scissor, short-axis CC$_{\text{ip}}$ stretch  \\
     &       & 1523 & 24.9     & 1577   & 57.6     &    \\                            
\hline
\end{tabular}
\end{table*}

The IR spectrum of cationic pyrene (\ce{C14H10+}, $m/z=202$ amu) is shown in Figure \ref{fig:pyr}a (black curve, top panel). It spans the range of 110--1590 \cm\ and consists of a composite of three individual curves, which overlap between 240--300 \cm\ and 560--600 \cm. We observe twelve distinct bands, for which frequencies and intensities are listed in Table \ref{tab:pyr}. The bands at 1320 and 1381 \cm\ have almost merged, but two distinct shapes can still be discerned. The observed bandwidths vary between approximately 40 and 60 \cm\ in the range between 450 and 1590 \cm, decrease to 30 \cm\ for the features at 335 and 434 \cm, and drop sharply to 3 \cm\ for the feature at 188 \cm.  
The spectrum between 1000--1600 cm$^{-1}$ is possibly affected by saturation due to the high power and strong absorptions; as a consequence, laser power normalisation leads to to an underestimation of the relative intensity. Although the high power may thus lead to saturation, it also enables the detection of very weak modes, especially in the FIR.
A vertical close-up (black, dotted curve) in the FIR section clearly reveals modes at 335 and 434 \cm\ with a high signal-to-noise (S/N) ratio, and predicted intensities of only 1.2 and 0.5 km$\cdot$mol$^{-1}$, respectively. These are around two orders of magnitude weaker than the strongest band in the MIR experimental spectrum. 

The 600--1590 cm$^{-1}$ region of our spectrum agrees well with a previously reported spectrum of \citet{Oomens2000} in terms of peak positions and band shapes. We detect an additional band at 1100 cm$^{-1}$ and confirm the band at 990 cm$^{-1}$, which was detected at a very low S/N ratio in the previous work. \citet{Panchagnula2020} also published an IRMPD spectrum of pyrene in their Supplementary Material, in the 600--1700 cm$^{-1}$ range,  which matches even better with our spectrum. They reproduced both our experimentally observed 990 and 1100 cm$^{-1}$ bands with a similarly high S/N, and the shape and peak positions line up nicely. In their main paper, they presented a high-resolution IR photodissociation spectrum of cold, neon-tagged pyrene in the same range. Many of the narrow bands in that spectrum  fall under one, broadened feature in their IRMPD spectrum. Notably, the left shoulders on the 840, 990, and 1230 cm$^{-1}$ bands in our spectrum can be explained by blending of the higher-intensity transitions with lower-intensity features. Furthermore, Panchagnula \textit{et al.} report a band at 768 \cm\, where we do not observe any fragmentation. 
The agreement is good, but this work exhibits a consistent redshift in the region of 1200--1600 \cm\ compared to the Ne-tagged spectrum, to a maximum of 23 \cm. This redshift may be attributed to the high laser power used in this work. 

Figure \ref{fig:pyr}a also displays the scaled harmonic (red curve, middle panel) and the anharmonic (blue curve, bottom panel) DFT spectra. 
The relative intensities for both theoretical spectra match the experiments reasonably well, although there are notable deviations in intensity for the observed 671, 840 and 1515 \cm\ bands. Most band frequencies in the scaled harmonic spectrum match within 10 cm$^{-1}$, except for the two experimental features at 990 and 1100 cm$^{-1}$. Here, the scaled harmonic positions at 953/968 cm$^{-1}$ and 1077 cm$^{-1}$ deviate 30--40 cm$^{-1}$ from experiment (see Table \ref{tab:pyr}). These two bands are predicted better in the anharmonic calculation, which only deviate by 8 and 9 \cm, respectively. 

Due to the sparse nature that the experimental and theoretical spectra exhibit in the FIR, unambiguous assignments are possible. Similar to the phenanthrene spectrum, the predicted FIR bands show activity in the entire skeletal frame, for which the exact descriptions are given in Table \ref{tab:pyr}. Above 671 cm$^{-1}$, the modes become increasingly local, and the experimental features are attributed to several close-lying local modes. The MIR assignments in Table \ref{tab:pyr} are therefore only given for the highest-intensity mode under the convolved peak. 

Arguably, the weakest band is found at 434 cm$^{-1}$, a band that does not seem to be predicted in either of the DFT spectra. However, a close-up of this wavelength region in Figure \ref{fig:pyr}b shows that the strong, predicted band at 485/500 cm$^{-1}$ (harmonic/anharmonic) is accompanied by two much weaker satellites. The low-frequency satellite could very well account for the experimental 434 cm$^{-1}$ band, while the high-frequency satellite could account for the asymmetric lineshape of the experimental 488 cm$^{-1}$ band. Including anharmonicity does not provide a better match with the experiment. The calculated intensity of the harmonic 451 cm$^{-1}$ band has an intensity of only 0.5 km$\cdot$mol$^{-1}$. Our ability to detect this feature again demonstrates the high experimental sensitivity. 

\subsection{Perylene}
\begin{figure*}
\includegraphics[width=\textwidth]{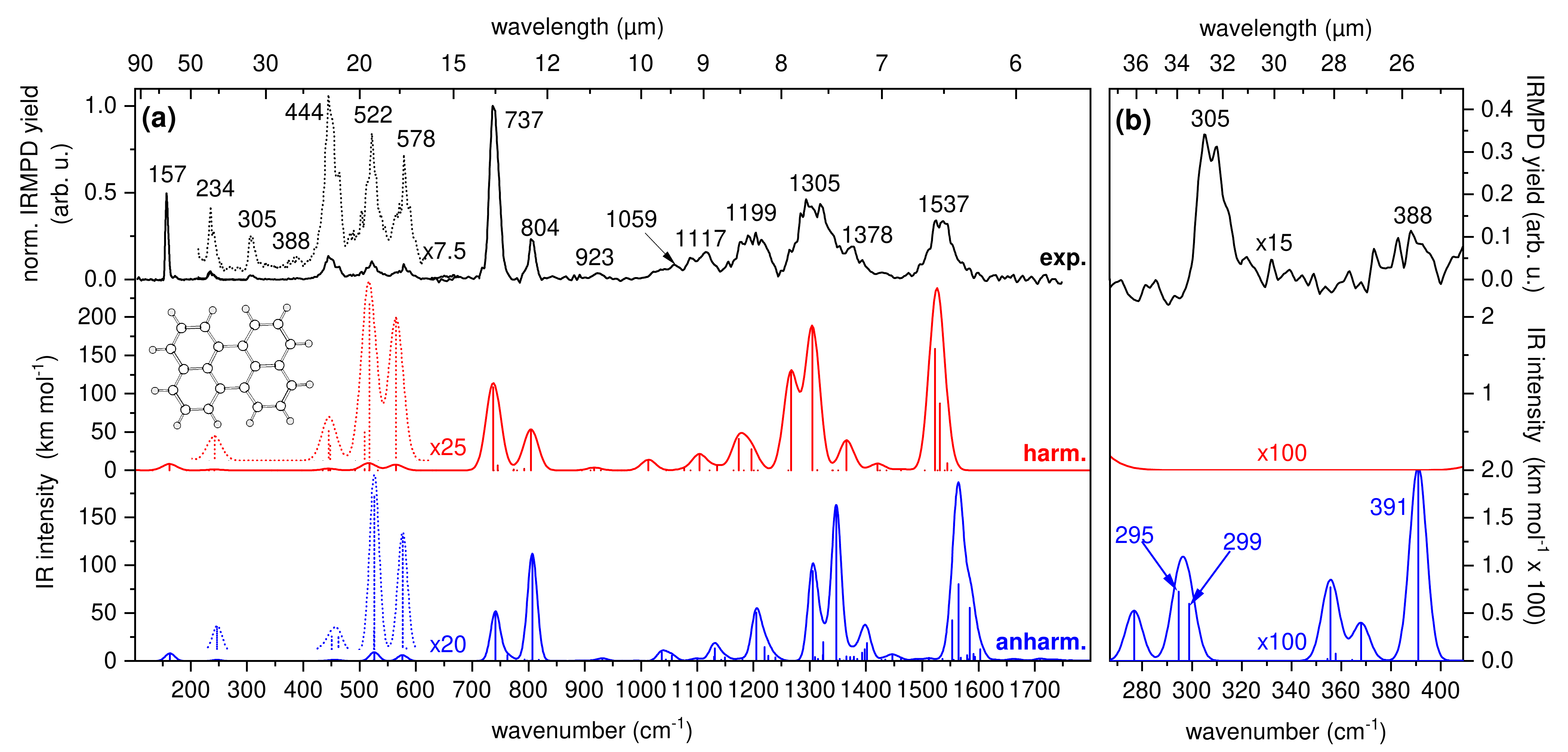}
\caption{\label{fig:per} 
a) The experimental IR spectrum of cationic perylene (top, black), accompanied by DFT calculated spectra in the harmonic (middle/red, frequencies scaled by a factor 0.96) and anharmonic (bottom/blue, unscaled) approximations; both calculated spectra are convolved with a 30 cm$^{-1}$ FWHM Gaussian line shape function. The dotted curves are vertical enlargements of the solid curves. b) Close-up of 270--410 \cm\ spectral range of panel a) Note that the values on the y-axis of the two bottom panels have been multiplied by 100 for increased readability, and that the convolution has been decreased to 10 \cm.
}
\end{figure*}

\begin{table*}
\caption{\label{tab:per} Perylene cation observed band frequencies (\cm) and relative intensities, together with calculated frequencies (\cm) and IR intensities (\km), and descriptions of the theoretical modes. The listed harmonic frequencies are scaled by a factor 0.96. \textit{oop: out-of-plane; ip: in-plane; E/C: elongation/compression}}
\begin{tabular}{ccccccl}
\hline
\multicolumn{2}{c}{Experiment}& 
\multicolumn{2}{c}{Harmonic calculations}&
\multicolumn{2}{c}{Anharmonic calculations}&\multicolumn{1}{l}{}\\
Freq. & Norm. $Y$ & Freq. & IR int. & Freq. & IR int. & \multicolumn{1}{l}{Mode description} \\
(cm$^{-1}$) &  & (cm$^{-1}$) & (km$\cdot$mol$^{-1}$) & (cm$^{-1}$) & (km$\cdot$mol$^{-1}$) & \multicolumn{1}{l}{} \\
\hline
157  & 0.495 & 162  & 8.4      & 163    & 8.0   & drumhead \\
234  & 0.044 & 242  & 1.3      & 247    & 1.3   & short-axis jumping jack \\
     &       &      &          & 295    & 0.007 & combination  \\
305  & 0.024 &      &          & 299    & 0.006 & combination \\
388  & 0.001 &      &          & 391    & 0.024 & combination \\
444  & 0.134 & 444  & 1.5      & 450    & 0.8   & long-axis jumping jack   \\
     &       & 447  & 0.8      & 462    & 0.8   &     \\
     &       & 509  & 1.5      & 525    & 1.1   &      \\
522  & 0.104 & 517  & 8.2      & 526    & 8.2   & zigzag CH$_{\text{oop}}$ wag \\
578  & 0.087 & 564  & 7.5      & 576    & 6.1   & long-axis E/C \\
737  & 1.000 & 737  & 108.8    & 741    & 51.5  & trio CH$_{\text{oop}}$ wag   \\
     &       & 746  & 6.5      & 762    & 6.1   &     \\
     &       & 773  & 1.2      & 778    & 0.6   &    \\
     &       & 793  & 2.1      & 793    & 1.5   &     \\
804  & 0.232 & 805  & 52.1     & 807    & 110.2 & armchair CH$_{\text{oop}}$ wag  \\
923  & 0.035 & 917  & 3.6      & 931    & 2.5   & antiphase zigzag CH$_{\text{oop}}$, bay CH$_{\text{oop}}$ \\
     &       & 1013 & 13.8     & 1037   & 9.1   & short-axis E/C          \\
1059 & 0.085 & 1077 & 1.8      & 1055   & 5.9   & trio CH$_{\text{ip}}$ scissor  \\
1117 & 0.157 & 1104 & 20.9     & 1131   & 13.2  & zigzag CH$_{\text{ip}}$ rock                 \\
     &       & 1135 & 5.6      & 1149   & 3.9   &          \\
     &       & 1174 & 41.0     & 1205   & 50.7  & antiphase armchair CH$_{\text{ip}}$ scissor\\
1199 & 0.270 & 1197 & 27.5     & 1220   & 14.5  & in-phase zigzag CH$_{\text{ip}}$ rock  \\
     &       & 1267 & 128.3    & 1305   & 93.9  & antiphase bay HCCCCH$_{\text{ip}}$ scissor  \\
     &       &      &          & 1324   & 19.8  &  \\
1305 & 0.462 & 1305 & 187.4    & 1347   & 158.4 & CC$_{\text{ip}}$ stretch \\
1378 & 0.189 & 1365 & 39.4     & 1393   & 8.7   & armchair CH$_{\text{ip}}$ rock \\
     &       & 1421 & 8.7      & 1398   & 12.5  &        \\
     &       &      &          & 1402   & 18.8  &           \\
     &       & 1523 & 158.8    & 1553   & 42.6  & zigzag CC$_{\text{ip}}$ stretch, zigzag HCCCH$_{\text{ip}}$ scissor \\
1537 & 0.341 & 1531 & 87.4     & 1564.6 & 78.4  & antiphase armchair CC$_{\text{ip}}$ stretch    \\
     &       & 1544 & 9.5      & 1565.0 & 80.4  &    \\
     &       &      &          & 1584   & 55.7  &     \\
     &       &      &          & 1603   & 12.5  &     \\                              
\hline
\end{tabular}
\end{table*}

For perylene (\ce{C20H12+}, $m/z=252$ amu), the IR spectrum in the 105--1760 cm$^{-1}$ range is shown in Fig. \ref{fig:per}a (black curve, top panel). The spectrum comprises three individual curves, with overlap between 215--300 \cm\, and 620--675 \cm. In total, we observe sixteen bands (see Table \ref{tab:per}) of which twelve can be clearly distinguished. The dotted curve shows a vertical enlargement for the weaker bands observed in the FIR, and clearly reveals high S/N bands at  234, 305, 388, 444, 522, and 578 cm$^{-1}$. The weakest 388 \cm\ feature has a relative intensity three orders of magnitude lower than that for the highest-intensity observed band at 737 \cm. Bandwidths decrease for lower frequencies, down to 5 \cm\ for the 188 \cm\ band.
The 188 \cm\ band has a line width of 5 \cm. A previously reported spectrum of perylene for the MIR region \citep{Bouwman2019}, agrees well with our spectrum in that range.  Band frequencies observed here are slightly redshifted (below 10 \cm) compared to \citet{Bouwman2019}. Additionally,  our work reveals two weak bands at 923 and 1059 cm$^{-1}$. 

Scaled harmonic (red curve, middle panel) and anharmonic (blue/bottom) spectra are also shown in Fig. \ref{fig:per}a. The harmonic prediction agrees well with the experimental curve. Most peak positions are reproduced within 10 cm$^{-1}$, although in the 1000--1550 cm$^{-1}$ range, deviations of up to 18 cm$^{-1}$ occur, potentially due to saturation effects. Deviations are larger for the anharmonic spectrum, although the overall shape agrees well. Following the same procedure as for phenanthrene and pyrene, the modes of most bands in the spectrum were assigned using the scaled harmonic spectrum, as listed in Table \ref{tab:per}. The sparse FIR spectrum allowed for unambiguous assignment, whereas the more densely populated MIR shows several calculated transitions present under one merged band. In the MIR, the assignment corresponds with the highest-intensity mode(s) under such a peak. Similar to pyrene, the modes in the 600--1600 cm$^{-1}$ range have an increasing in-plane component for increasing frequency. In the FIR (100--600 cm$^{-1}$) region, we find skeletal out-of-plane and elongation/compression (E/C) modes. 

The harmonic spectrum shows no intensity in the range between 300--400 \cm, where two features are observed experimentally at 305 and 388 \cm. Figure \ref{fig:per}b displays a deeper close-up of this range. The scaled harmonic curve only shows activity at the very edges. The anharmonic prediction shows minor activity, resulting from combination bands --- not exceeding intensities above 0.02 \km. 
The 305 \cm\ band is assigned to two combination bands calculated at 295 \cm\ (the 133 \cm\ \ce{H-C-C-H} twist/scissor vibration and the 163
\cm\ drumhead mode) and at 299 \cm\ (two ring twisting modes at 26 and 274 \cm), whereas that at 388 \cm\ is assigned to a combination band calculated at 291 \cm\ (the 26 \cm\ ring twisting
mode and a 364 \cm\ diagonal E/C mode).
The experimental 388 cm$^{-1}$ feature is assigned to a combination band predicted at 391 cm$^{-1}$ with an intensity of 0.024 km$\cdot$mol$^{-1}$, which is a combination of an IR-inactive ring twisting mode at 26 cm$^{-1}$ and an IR-inactive diagonal E/C mode, at 364 cm$^{-1}$. Calculation predicts an opposite intensity ratio between the two features observed in experiment. It may very well be that the closeness of the 295 and 299 cm$^{-1}$ bands leads to cooperative absorption, resulting in one stronger resonance.

\section{Discussion}

\begin{figure}
\centering
\includegraphics[width=\linewidth]{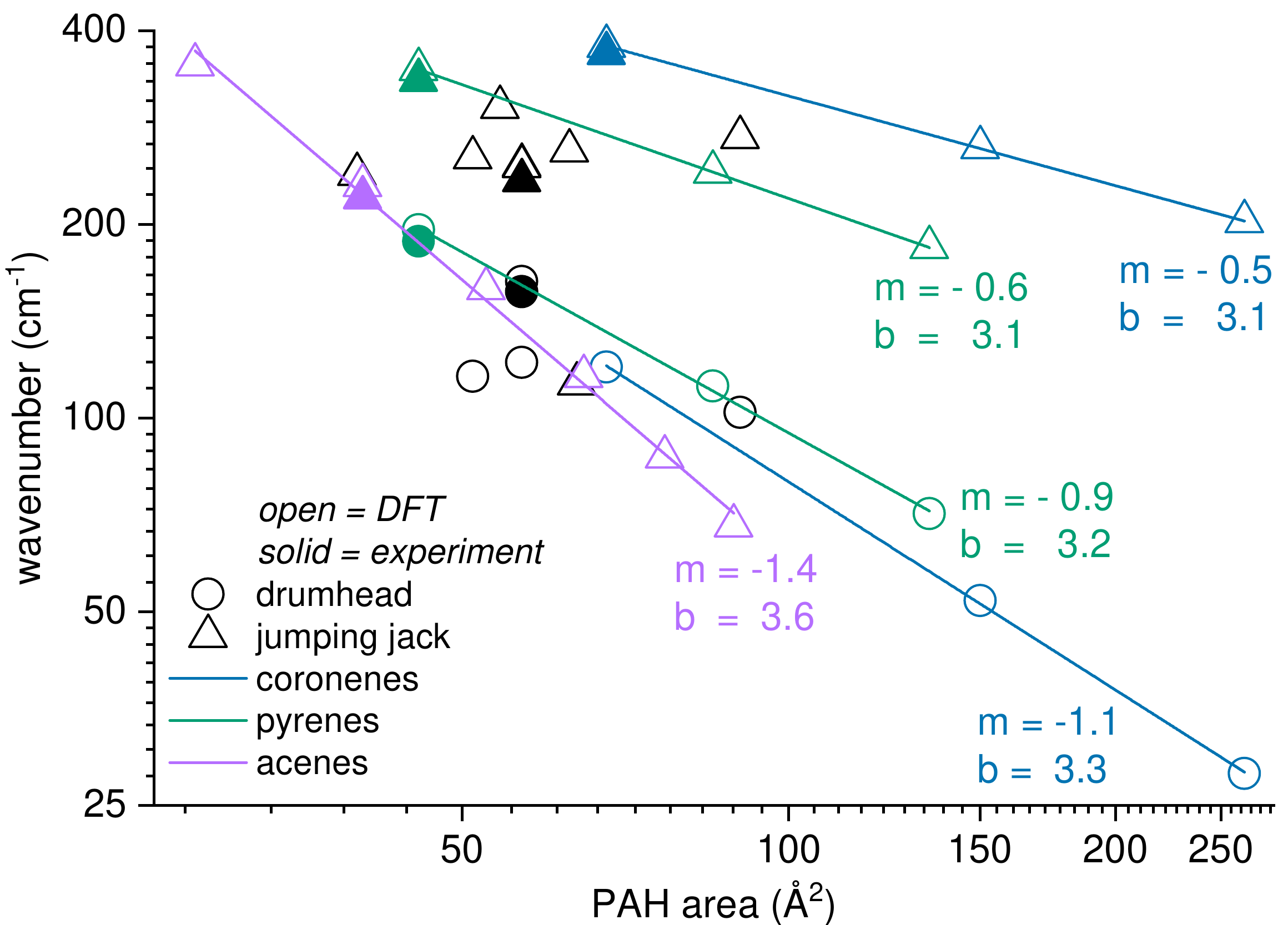}
\caption{\label{fig:NASA} Double logarithmic plot of the harmonic DFT calculated (0,1) drumhead modes (circles) and short-axis jumping jack (triangles) for twenty cationic PAHs present in the NASA Ames PAH Database (PAHdb) \citep{Bauschlicher2018}, together with experimentally observed band positions of four of these molecules as reported previously and in this work (open symbols) (anthracene and coronene from \citet{Bakker2010}, and pyrene and perylene from the present study). Linear fits of the predicted bands as function of PAH surface area for the different molecular families are given by the solid lines. Between brackets are the values of m and b for the fit as given in Eqn. \ref{eqn:2}.  A full list of included species is given in Table \ref{tab:NASA}. Cationic perylene was not present in the NASA Ames database, and we therefore used our own harmonic calculation, scaled with a factor of 0.96 to match in both the MIR and the FIR.}
\end{figure}

\begin{table}
\caption{\label{tab:NASA} List of PAH cations plotted in Figure \ref{fig:NASA}, with their size in number of rings $R$ and their area $A$ in \AA$^{2}$. In the abbreviated nomenclature, capital B stands for `benzo-', capital DB for `dibenzo-' and capital C, for `circum-'.}
\begin{tabular}{lcclcc}
\hline
\multicolumn{3}{c}{coronenes}            & \multicolumn{3}{c}{pyrenes}               \\
name                 & $R$ & $A$ (\AA$^{2}$) & name                  & $R$ & $A$ (\AA$^{2}$) \\ \hline
coronene             & 7     & 68        & pyrene                & 4     & 46         \\
Ccoronene            & 19    & 150       & DB{[}bc,kl{]}coronene & 9     & 85        \\
CCcoronene           & 37    & 263       & DB{[}bc,op{]}Cpyrene & 16    & 135        \\\hline
\multicolumn{1}{c}{} &       &           & \multicolumn{1}{c}{}  &       &           \\ \hline
\multicolumn{3}{c}{acenes}               & \multicolumn{3}{c}{differently-shaped}        \\
name                 & $R$ & $A$ (\AA$^{2}$) & name                  & $R$ & $A$ (\AA$^{2}$) \\ \hline
naphthalene          & 2     & 28        & phenanthrene          & 3     & 40         \\
anthracene           & 3     & 40        & triphenylene          & 4     & 51        \\
tetracene            & 4     & 53        & perylene              & 5     & 57        \\
pentacene            & 5     & 65        & B{[}e{]}pyrene        & 5     & 57        \\
hexacene             & 6     & 77        & olympicene            & 5     & 54        \\
heptacene            & 7     & 89        & DB{[}a,j{]}anthracene & 5     & 64        \\
\multicolumn{1}{c}{} &       &           & anthanthrene           & 6     & 63        \\ 
\multicolumn{1}{c}{} &       &           & ovalene               & 10    & 90         \\ \hline
\end{tabular}
\end{table}

The MIR-to-FIR spectra presented in this work are the first spectra of gas-phase, cationic PAHs to go down to 100 cm$^{-1}$, with bands below 200 \cm\ having bandwidths of $\leq 5$ \cm. These spectra represent an important venture into the FIR, adding to very limited experimental data \citep{Bakker2010}.
Astronomical FIR spectroscopy has been proposed in various studies as a method for molecule-specific PAH identification in the ISM \citep{Boersma2011a,Boersma2011b,Mulas2006a,Mulas2006b,Ricca2010,Ricca2012}. Such studies ended up largely fruitless due to the lack of astronomical sensitivity at the time. With the  Origins Space Telescope (OST) mission on the horizon, we are hopeful that efforts to find FIR signatures will be reinvigorated. With its high-resolution OSS and HERO FIR instruments, the OST mission will provide a resolution and sensitivity that are orders of magnitude better than Herschel's PAC and AKARI's FIS instruments were able to offer \citep{Battersby2018}. This mission thus offers a realistic (and exciting) prospect 
to discern PAH FIR modes from the dust background. With this increased accuracy, the demands on the accuracy of the theoretical predictions will be increased as well. Considering that all FIR calculations have been benchmarked on experimental values of neutral naphthalene \citep{Pirali2006,Pirali2009}, large improvements are possible if a larger set of experimental spectra is used for benchmarking. 

We follow the works of the NASA Ames group  \citep{Ricca2010,Ricca2012,Boersma2011a,Boersma2011b}, and present FIR frequency trends for cationic PAHs as a function of PAH shape and size, using the currently available IRMPD data and harmonic DFT calculations. In Fig. \ref{fig:NASA}, experimental values (open symbols) are compared to DFT calculated values (solid symbols), as a function of PAH surface area $A$ (defined in \citet{Ricca2012}). The experimental data are from the current work (pyrene and perylene) and from a previous study (anthracene and coronene, \cite{Bakker2010}); the calculated values from the NASA Ames PAH Database (PAHdb, \cite{Bauschlicher2018}), comprising a larger subset of PAHs. To enhance readability, only the drumhead modes with a single node (the (0,1) modes, circles), and short-axis jumping jack modes (triangles) are plotted. The PAHs are classified in families that are color coded. Names, number of rings, surface area and family classification for each species are detailed in Table \ref{tab:NASA}.

If we compare the experimental to theoretical values, it is clear that they match very well. On closer inspection, it can be seen that they exhibit a systematic redshift of 6--11 cm$^{-1}$. It is not directly clear what causes this redshift, but two potential causes are the finite temperatures at which experiments are conducted, and the intrinsic red-shifting of IRMPD bands compared to the fundamental $\nu=1 \leftarrow \nu=0$ transitions that are calculated.

To investigate trends, the calculated frequencies for one vibration type are fitted to a linear function of PAH surface area $A$, for several families:
\begin{equation}\label{eqn:2}
    \log(\Omega)=\log(mA+b) \rightarrow \Omega=10^{b}A^{m}+C,
\end{equation}

\noindent with $\Omega$ the frequency (in cm$^{-1}$) and $A$ the PAH area (in \AA$^{2}$). The resulting coefficients $m$ and $b$ are given in Fig. \ref{fig:NASA}.

We discern a clear area dependence of the frequencies of jumping jack modes for all families. The strongest dependence is found for the acene family (purple triangles). This strong size-dependence closely follows what was identified by \citet{Boersma2011a} for the bar modes (out-of-plane folding vibrations) of acenes. Their jumping jack modes can also be considered to be folding vibrations, although in-plane.
{Two of the mixed-shape PAHs, phenanthrene and dibenzo[a,j]anthracene, appear to coincide with the observed acene trend line, likely because these PAHs also consist of only a single row of rings.} The jumping jack frequencies for the families of coronenes (blue triangles) and the pyrenes (green triangles) exhibit a significantly weaker size-dependence. Ricca and co-workers identified similar trends for jumping jack modes of very large, neutral PAHs \citep{Ricca2010}, and of coronene-like PAHs for which they predicted a convergence at $\sim$200 cm$^{-1}$ \citep{Ricca2012}. This result is reflected in the relatively weak size-dependence of the jumping jack frequencies for the coronene family. Given the similarity of the size-dependence here identified for the pyrene family, we propose a similar convergence or ‘pile-up’, albeit at lower frequency could occur for them.

The absolute frequencies of drumhead modes for both the coronenes (blue circles) and the pyrenes (green circles) are lower than those of the jumping jack modes, but they exhibit stronger size-dependencies, reflected in the steepness of the fits. Given the similarities between both families the shape dependence is rather weak. The strong size-dependence was previously found by \citet{Boersma2011b}. Again, two mixed-shape PAHs, perylene and ovalene, follow the trend observed, potentially because both they and pyrenes in general have \ce{D_{2h}} symmetry. Because the drumhead trends for the coronenes and pyrenes show a weak, but significant difference (at $A=200$ \AA$^{2}$ it would be $\sim7$ \cm\ $\pm$ constant $C$, see Eq. \ref{eqn:2}), drumhead modes are slightly, but significantly shape dependent, contrary to what was previously postulated. Triphenylene and benzo[e]pyrene, two other two mixed-shape molecules with a drumhead mode, do not coincide with the derived trend lines.

The trends identified here are potentially helpful in the identification of unique PAHs or PAH families using current and future observational instruments. The frequency convergence identified for jumping jack modes of coronenes and pyrenes could help identify them as family, whereas drumhead modes would be markers for individual PAHs. However, it remains crucial that the theory upon which they are based is benchmarked with accurate experimental data, especially for larger PAHs. Given the current demonstration of IRMPD spectroscopy extended toward 100 $\upmu$m, we foresee that such data will become available in the coming years.

\section{Summary and Conclusion}
We present gas-phase spectra of cationic PAHs in the MIR to FIR region, up (down) to 100 $\upmu$m (100 cm$^{-1}$). These spectra match our scaled, harmonic spectra very well.  We also present anharmonic DFT calculations for cationic PAHs in the FIR, which we used to assign pure combination bands with a predicted intensity down to 0.006 km$\cdot$mol$^{-1}$. The lowest-frequency bands also match calculations from the NASA Ames PAH Database \citep{Bauschlicher2018} fairly well, although a small, systematic redshift is found. The bandwidths below 200 \cm\ are $\leq5$ \cm, making the IRMPD data in this wavelength range suitable for benchmarking efforts. 

In this context, we revisited the hypothesis that FIR bands can be used to search for the IR emissions of single PAH species \citep{Mulas2006a,Mulas2006b,Boersma2011a,Boersma2011b,Ricca2012}. We report a larger dependence on molecular shape than was previously found, but emphasise the importance of reliable benchmarking before the deduced trends are used to search for astronomical features. Future experimental efforts should be focussed on the measurement of FIR spectra of large, symmetric PAHs, both neutral and cationic. 
The future Origins Space Telescope will be able to provide the sensitive observational data that is necessary to find these elusive FIR bands \citep{Joblin2011,Battersby2018}.

\section*{Acknowledgements}
We thank Dr. Christine Joblin for the insightful discussions and Prof. W.J. Buma for his thorough reading of an earlier version of the manuscript. We thank COST Action CA18212 - Molecular Dynamics in the GAS phase (MD-GAS), supported by COST (European Cooperation in Science and Technology). We gratefully acknowledge the {\it Nederlandse Organisatie voor Wetenschappelijk Onderzoek} (NWO) for the support of the FELIX Laboratory. This work is supported by the VIDI grant (723.014.007) of A.P. from NWO. Furthermore, A.C. gratefully acknowledges NWO for a VENI grant (639.041.543). Calculations were carried out on the Dutch national e-infrastructure (Cartesius and LISA) with the support of Surfsara, under projects NWO Rekentijd 16260 and 17603.

\section*{Data Availability}
 
The data underlying this article is available at \url{www.doi.org/10.21942/uva.17023880}. Theoretical data from Figure \ref{fig:NASA} is available at \url{https://www.astrochemistry.org/pahdb/theoretical/3.20/default/view}, and several experimental data points are derived from \citet{Bakker2010}.



\bibliographystyle{mnras}
\bibliography{mnras_p3} 








\bsp	
\label{lastpage}
\end{document}